\documentstyle[aps,preprint]{revtex}
\begin{document}
\draft
\title{Sandpile Model with Activity Inhibition}
\author{ S. S. Manna$^1$\nocite{add1}  and D. Giri$^2$\nocite{add2} }
\address{ Department of Physics, Indian Institute of Technology \\
     Powai, Mumbai 400076, India }
\date{\today}
\maketitle
\begin{abstract}
A new sandpile model is studied in which bonds of the system are
inhibited for activity after a certain number of transmission of 
grains. This condition impels an unstable sand column to distribute 
grains only to those neighbours which have toppled less than $m$ times. 
In this non-Abelian model grains effectively move faster than 
the ordinary diffusion (super-diffusion). A novel system size 
dependent cross-over from Abelian sandpile behaviour to a new 
critical behaviour is observed for all values of the parameter $m$.
\end{abstract}

\narrowtext

%%%%%%%%%%%%%%% TEXT STARTS HERE %%%%%%%%%%%%%%%%%%%%%%%%%%%%%%%%%%

The concept of Self-Organized Criticality (SOC) was introduced
to describe how a system, starting from an arbitrary initial condition
may evolve to a scale free critical state following some specific 
dynamical rules while under the action of repeated external 
perturbations [1]. Naturally occurring physical phenomena like
sandpiles [2], forest fires [3], river networks [4], earthquakes 
[5] etc. are argued as systems showing SOC. To demonstrate the idea of 
SOC a simple model known as the `sandpile' model was introduced in 
which a stochastically driven cellular automata evolves under a 
non-linear, diffusive, self-organizing mechanism leading to a 
non-equilibrium critical state [1].

At present many different versions of the sandpile model are available. 
However precise classification of various models in different 
universality classes in terms of their critical exponents is not yet 
fully complete and still attracts much attention [6]. Among the 
different models most widely studied is the Abelian Sandpile Model 
(ASM) in which many analytical [7] as well as numerical [8] results 
are known. Some efforts have also been given towards the analytical
calculation of avalanche size exponents [9,10].
Secondly a Two-state sandpile model with 
stochastic evolution rules was also studied [11] which was initially 
thought to belong to the same universality class as that of ASM 
[11,12] but later claimed to be different [13].

We consider a situation in which an intermediate time scale
is associated with every bond of the system. Each bond allows only 
a certain number of grains to cross from its one end to the other and
after that it has a dead time and cannot support any further
traffic untill a new avalanche starts. This dead time is much greater than
the time scale of avalanche propagations but much less than the 
input rate of grains. We call this model as the `Sandpile
Model with Activity Inhibition' (SMAI).

Similar to different sandpile models we also define our model on a 
regular lattice with open boundary. Non-negative integer numbers 
($h_i$) assigned at the 
lattice sites represent the heights of the sand columns. Sand grains 
are added at randomly chosen sites by increasing the $h$ values by 
unity : $h_i \rightarrow h_i + 1$. The possibility of a sand column 
becoming unstable arises only when the height $h_i$ becomes greater 
than a threshold value $h_c$. Such a column becomes unstable 
only if the number $n_i$ of nearest neighbour sites which have toppled 
less than a pre-assigned cut-off number $m$ within the same avalanche is found 
to be non zero. An unstable column immediately topples and distributes
one grain each to all the $n_i$ neighbours :
$h_j \rightarrow h_j + 1$ ($ j=1$ to $n_i$).
The sand column decreases by the same amount : $h_i \rightarrow h_i - n_i$.
If $n_i = 0$, the sand column does not topple and its height
though greater than $h_c$ is considered stable.
In an avalanche sites can topple a maximum of $m$ times.
This implies that in the limit of $m \rightarrow \infty$
our model converges to ASM.
Recently a Stochastic Sandpile Model (SSM) has been studied
in which sand columns having heights greater than the threshold
are also considered stable [14].

One unit of time within an avalanche consists of the following
intermediate steps : (i) a list of all sites where $h_i > h_c$ is
made (ii) $n_i$ values are calculated for each site $i$
(iii) all sites with non-zero $n_i$ values are toppled in parallel.

We first consider the case where the cut-off in the toppling 
number $m$ = 1. Here the toppling front moves outwards
and grains always jump only in the outward direction
and do not fall back.
Therefore compared to the random walk analogy for the movement of the
grains in ASM [15] in our model grains move
faster than diffusing particles. This is indeed reflected in
the average cluster size $<s> \sim L^{\beta_s}$ where 
$\beta_s = 1.62$ (reported below). This 
implies that the displacements ${\cal R}$ of the grains in our model
grows with time ${\cal T}$ as ${\cal R} \sim {\cal T}^\nu$ 
with $\nu = 1/1.62 = 0.62$
which is faster than diffusion (super-diffusion). 

Zhang had studied a scaling theory of the sandpile model
in which the toppling front grows as a $(d-1)$ dimensional
surface in the $d$ dimension [16]. In this analysis, where 
multiple topplings were ignored, the exponent for the
avalanche size distribution was possible to calculate 
analytically in all dimensions. Since in SMAI, a single 
toppling front moves outward and multiple topplings are
forbidden for $m$ = 1, we expect that SMAI
may be a correct realization of the Zhang's theory [16].

Unlike ASM our model turns out to be non-Abelian.
Different steady state configurations are obtained on
dropping grains at same locations but following different
sequences. We pick up a stable configuration in a
$2 \times 2$ cell and drop one grain each at the two
opposite corners one after the other (figure 1).
On reversing the order of dropping a different 
stable state configuration
is obtained. We notice that non-abelianity is effective only when
avalanche cluster sizes are greater than one.

However, it is easy to see that just like in ASM, we cannot 
have the forbidden subcofigurations (FSC) anywhere in the 
lattice. An FSC is defined as the subset of connected 
sites for which at each site the height is less than its 
coordination number in the subset [7]. In SMAI also two 
neighbouring sites whose heights are both zero ($0-0$) will 
never occur in the steady state because of the same reason 
as in ASM that if one topples the other site will receive 
one grain. Similarly a height configuration like $(0-1-0)$ 
is also an FSC. In fact, all the FSCs defined for ASM are 
also forbidden here, and a recurrent config must burn
completely. This is checked by burning many successive steady 
state configurations where the fire starts from the boundary 
as well as from those sites with heights greater than $h_c$ [7].
All the configurations are observed to burn fully. No unburnt
configuration is found which is obtained by adding grains to 
a steady state that burns.

To use the rotational symmetry of the system the sandpile is 
grown with $h_c = 3$ within a circular region of radius $R = 
(L-1)/2$ placed on a square lattice of size $L \times L$. In 
the steady state starting from the boundary the average height 
grows quickly radially towards the centre following a power law 
: $<h(r)> = A - B (R-r)^{-\delta}$, where $r$ is the radial distance 
measured from the centre. We estimate $A$ = 2.3904, $B$ = 7.81 and 
$\delta = 0.75$ for $L$ = 1025 (figure 2). The average height per site is 
found to depend on $L$ which on plotting with $1/L$ extrapolates 
to a value 2.3840 in the limit of $L \rightarrow \infty$. Similar 
analysis yields the fraction of sites with different column heights 
are approximately 2.1$\times 10^{-4}$ ($h=0$), 0.2421 ($h=1$), 
0.3059 ($h=2$), 0.3404 ($h=3$). Beyond $h=3$, this fraction 
decreases approximately exponentially as $\exp(-\alpha h)$ where 
$\alpha = 1.64 $ and adds up to a total of 0.1118.

The size of the avalanche is measured in three different 
ways : (i) the total number sites $s$ which crosses the threshold $h_c$
(either toppled or not, both counted) (ii) the life time of 
the avalanche ($t$) and (iii) the linear extent or the radius 
($r$) of the avalanche. 
Since $s, t$ and $r$ are the three different measurements of the
same random avalanche cluster, they are necessarily dependent
variables. These quantities are assumed to depend on one another as 
\begin {equation}
s \sim t^{\gamma_{ts}}, \quad \quad
r \sim t^{\gamma_{tr}}, \quad \quad
s \sim r^{\gamma_{rs}}. \quad \quad
\end {equation}
The three $\gamma$ exponents are connected by the relation
\begin {equation}
\gamma_{ts} = \gamma_{tr} \gamma_{rs}
\end {equation}

To estimate the exponents $\gamma_{ts}$ and $\gamma_{tr}$ we
measure the avalanche size $s$ and avalanche radius $r$ at 
every time step $t$ during the progress of each avalanche.
The total number of topplings up to time $t$ gives the
intermediate size $s$ where as the size of the smallest square which 
encloses the cluster gives the intermediate radius $r$. 
We estimate $\gamma_{ts} = 1.64 $ and
$\gamma_{tr} = 0.83 $.
Since the avalanche clusters are quite compact and has only few
small holes it is justified to assume that $\gamma_{rs} = 2$. 
These values very closely satisfy the equation (2).

We assume the finite size scaling forms 
for the probability distribution functions as
\begin {equation}
P(s) \sim s^{-\tau_s} f_s \left (\frac {s}{L^{\sigma_s}}\right ), \quad
P(t) \sim t^{-\tau_t} f_t \left (\frac {t}{L^{\sigma_t}}\right ), \quad
P(r) \sim r^{-\tau_r} f_r \left (\frac {r}{L^{\sigma_r}}\right ). \quad
\end {equation}
Consequently the cumulative probability distribution 
$F(x) = $$\int^{L^{\sigma_x}}_x P(x)dx $ varies as $x^{1-\tau_x}$.
However, in the case of $\tau_x = 1$, the variation should
be in the form $F(x) = C -$ log$(x)$.

We plot the data of $F(s)$ in two different ways. In figure 3 
we plot $F(s)$ vs. $s$ for system sizes $L =$ 65, 257 and 1025
using a log $-$ lin scale. Presence of humps in the large $s$ limit
is visible for bigger system sizes
which reflects the effect of the finite system
size on power law distributions. However in the intermediate
region curves are reasonably straight indicating that
the exponent $\tau_s$ is likely to be 1. We further plot 
$F(s) s^{\tau_s(L) - 1}$ with $s$ on a log $-$ log scale
and tune $\tau_s(L)$, the effective $\tau_s$ exponent
for the system size $L$, such that the curves become horizontal
in the intermediate range of $s$. 
All three curves collapse nicely when the abscissa is
scaled as $s L^{-1.62}$ which implies that $\sigma_s = 1.62$.
We show in the figure 4
that the $\tau_s(L)$ values very closely fit to a straight line
when plotted with $L^{-1/4}$. It seems that $L^{-1/4}$
may be the right leading correction to scaling.
The fitted straight line when extrapolated
to $L \rightarrow \infty$ gives a value of 1.016 for $\tau_s$.
Similar analysis for the life time distribution also leads us to
conclude that $\tau_t = 1.02 $, $\sigma_t = 0.98$.

The radius distribution $F(r)$ is calculated in a slightly
different way. We estimate the probability that a site at a
distance $r$ from the centre of mass of the avalanche cluster
is a part of the cluster. We take into consideration the 
`degeneracy' effect that different sites could be at exactly
equal distances from the centre. In figure 5 we show a
scaling plot $F(r) L^{0.20}$ against $r L^{-0.86}$
using a log $-$ lin scale for different system sizes.
Here we see a much better straight part in the intermediate
region. We conclude a value of $\tau_r \approx 1$.

The distribution functions follow relations like
$P(s)ds \sim P(t)dt$ which imply following scaling relations
\begin {equation}
\tau_s -1 = \gamma_{ts} (\tau_t -1), \quad
\tau_r -1 = \gamma_{tr} (\tau_t -1), \quad
\tau_s -1 = \gamma_{rs} (\tau_r -1).
\end {equation}
These equations imply that if one of the exponents
$\tau_s, \tau_t$ or $\tau_r$ is equal to one, rest are
also equal to one, irrespective of the values of the
$\gamma$ exponents. Our estimates for the different $\tau$
exponents are very much consistent with these equations.
We also observe that the value of $\tau_s \approx 1$
agrees very well with the Zhang's result
$\tau_s = 2(1-1/d)$ for $d$ = 2 [16].

We also assume that the average values of
$s, t$ and $r$ varies with the system size $L$ as
\begin {equation}
<s(L)> \sim L^{\beta_s}, \quad \quad \quad
<t(L)> \sim L^{\beta_t}, \quad \quad \quad
<r(L)> \sim L^{\beta_r}. \quad \quad \quad
\end {equation}
We plot $<s(L)>$ vs. $L$ on log $-$ log scale
for $L$ = 33, 65, 129, 257, 513 and 1025. Slopes between
successive points are plotted with $L^{-2}$ and 
extrapolated to $L \rightarrow \infty$ limit giving
$\beta_s = 1.61$. Similar analysis gives
$\beta_t = 0.96$ and $\beta_r = 0.82$.

Using the scaling forms in equation (3) we get following 
scaling relations for $\beta$ exponents as
\begin {equation}
\beta_s = \sigma_s (2-\tau_s), \quad \quad \quad
\beta_t = \sigma_t (2-\tau_t), \quad \quad \quad
\beta_r = \sigma_r (2-\tau_r). \quad \quad \quad
\end {equation}
With our measured values of $\beta, \sigma,$ and $\tau$
these relations are approximately satisfied.
We put errors of 0.05 to all our measured exponents.

Next we study the case when cut-off for the toppling
number $m > 1$. The average cluster size $<s(L)>$ is
plotted with $L$ on the log $-$ log scale in figure 6
for $m$ = 1, 2, 4 and 8. We see that all curves are
parallel straight lines with slopes approximately 1.61
for large system sizes. However for small system 
sizes all of them bend and become part of the same straight
line. Then we plot on the same figure the $<s(L)>$ data for
ASM. We get a straight line with a slope $\approx$ 2
which almost overlaps
with the bend portions of the curves for different $m$ values.
We explain this by noting that for every $m$ value
our model behaves as ASM for small system sizes. 
In small systems the number of avalanches where
sites will topple more than $m$ times are very few.
However for bigger system sizes the cut-off $m$ will
have more prominent effects. Therefore for each $m$
values there should be one particular system size
where the cross-over takes place from ASM to Non-abelian
behaviour. The cross-over size $L_c$ is seen to be roughly
proportional to the value of $m$. We expect that for any 
$m$ value if one
works in systems larger than the cross-over size one
should get the same set of exponents as those in the case of $m$ = 1.

To summarize, we studied here a new sandpile model where
bonds of the system relax after a certain number of transmission
of grains. This limits a site to topple a maximum of $m$ times
within the same avalanche.
Based on the results of detailed numerical studies using
improved algorithms we claim a cross-over from ASM 
behaviour to a new critical behaviour 
at a particular size of the system whose magnitude depends
on the value of $m$.

We acknowledge with thanks D. Dhar for many useful discussions 
and suggestions.

\begin{figure}
\caption{Non-abelian property of the sandpile model
is shown on a $2 \times 2$ cell. On the same initial
stable configuration two grains are added at two different
sites but in different orders. Different final stable configurations
are obtained.}
\end{figure}

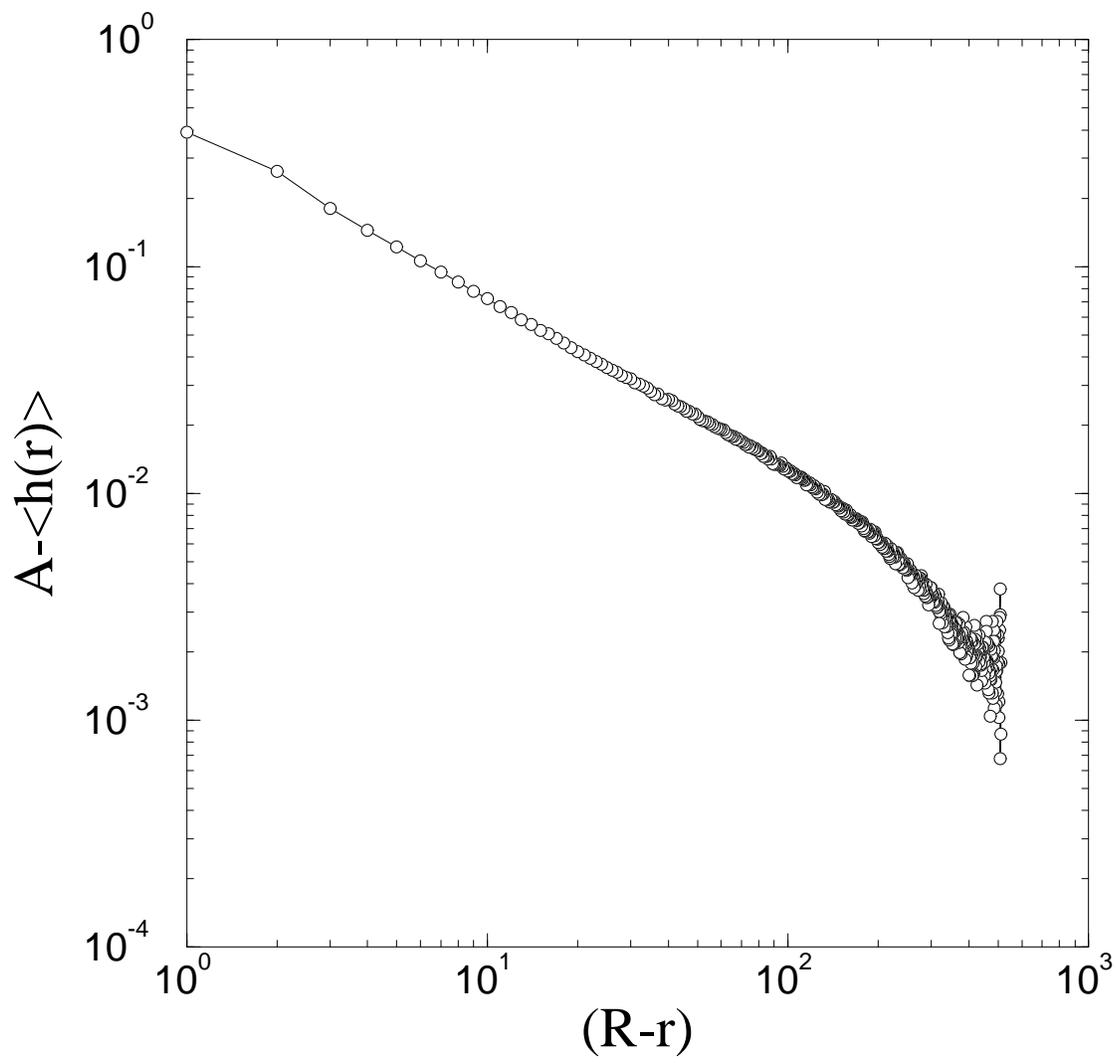
\begin{figure}
\caption{The average height profile of the sandpile
in a circular region plotted with the distance from the
circumference of the circle is shown. The slope of the
curve is 0.75 and $A$ = 2.3904 is found.}
\end{figure}

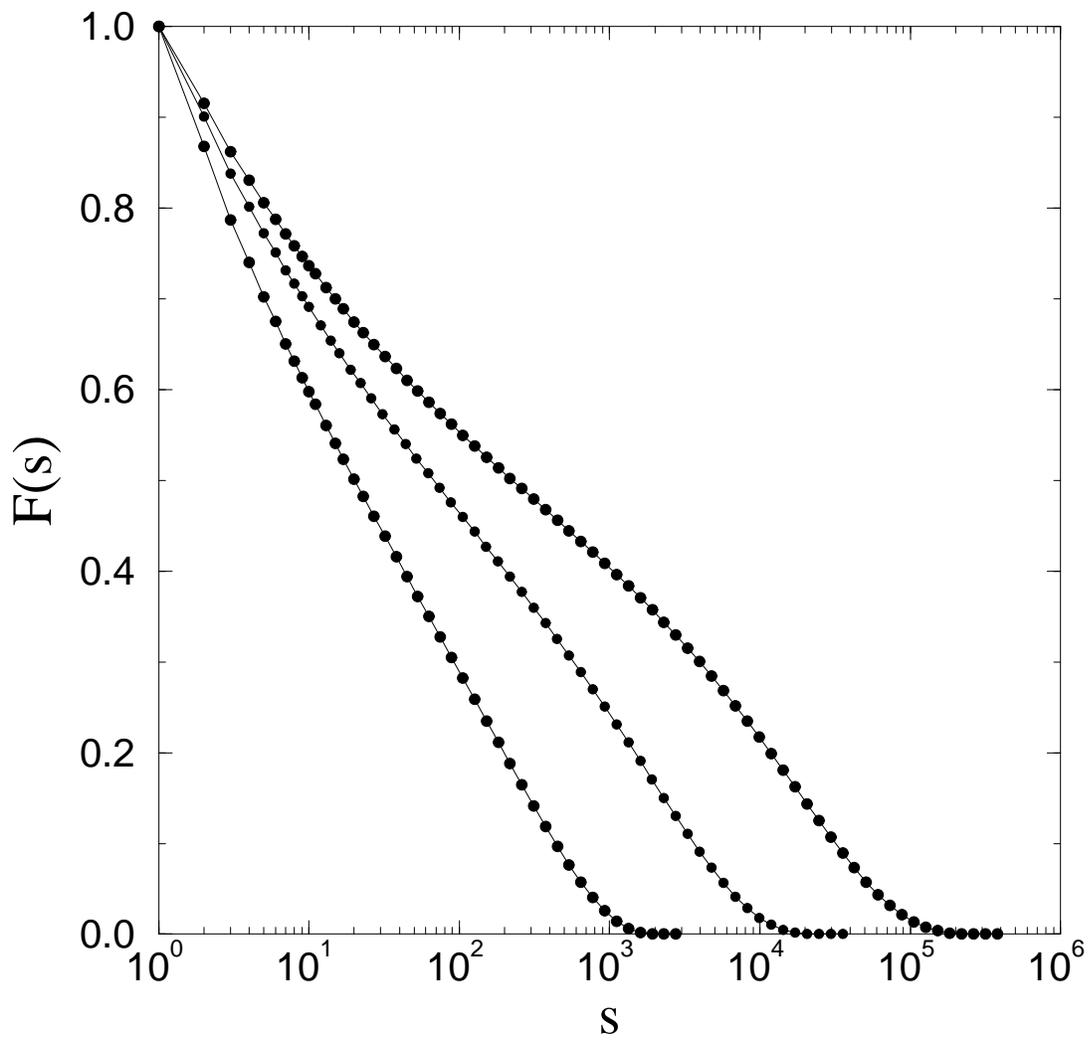
\begin{figure}
\caption{Log $-$ lin plot of the cumulative probability
distribution $F(s)$ for the three system sizes
$L = 65, 257$ and $1025$ (from left to right). 
The straight portions of the
curves in the intermediate regions indicates that
$\tau_s$ is likely to be equal to one.}
\end{figure}

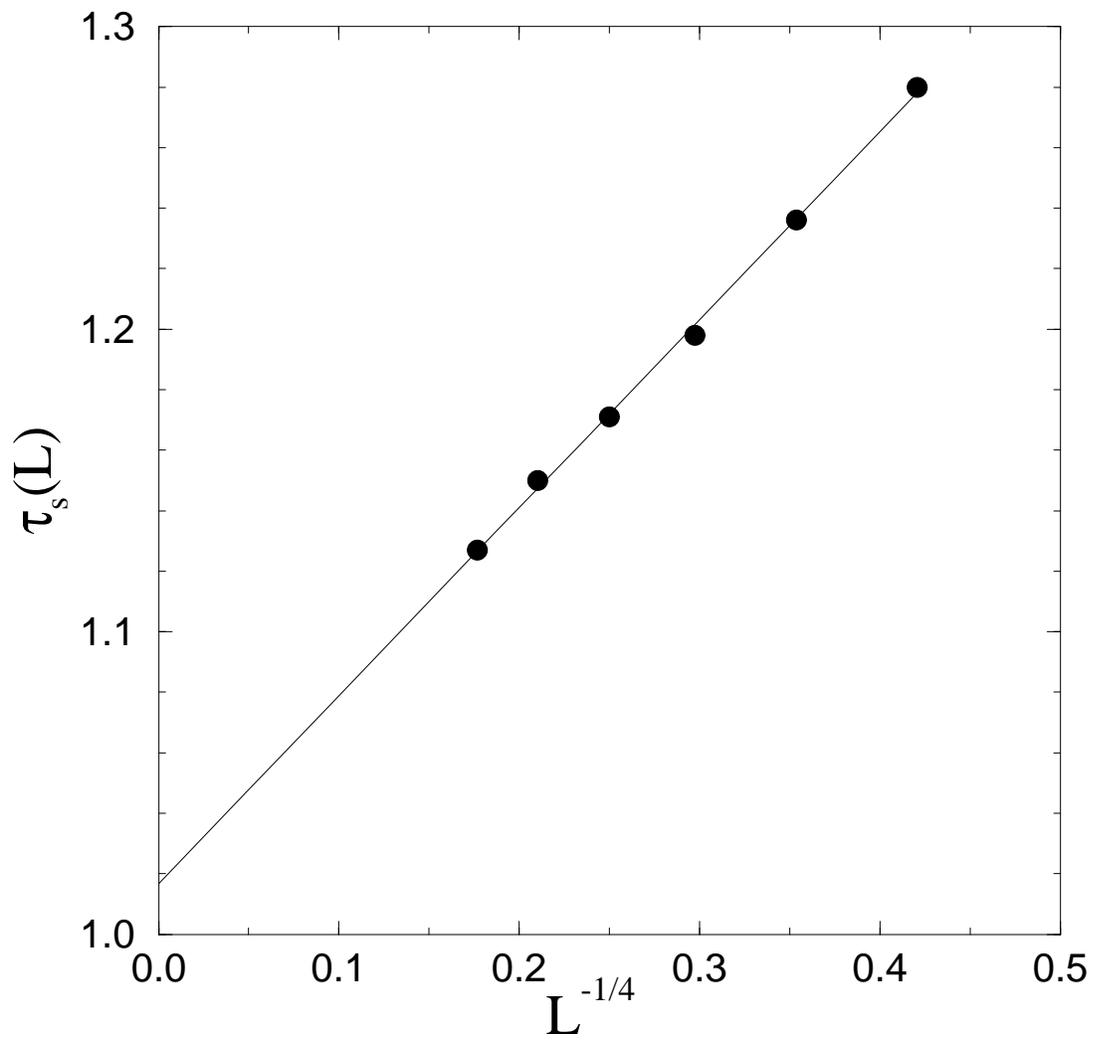
\begin{figure}
\caption{Plot of $\tau_s(L)$ for different system
sizes $L = 33, 65, 129, 257, 513$ and $1025$ with
$L^{-1/4}$. A direct straight line fit gives
$\tau_s$ = 1.016 in the $L \rightarrow \infty$ limit.}
\end{figure}

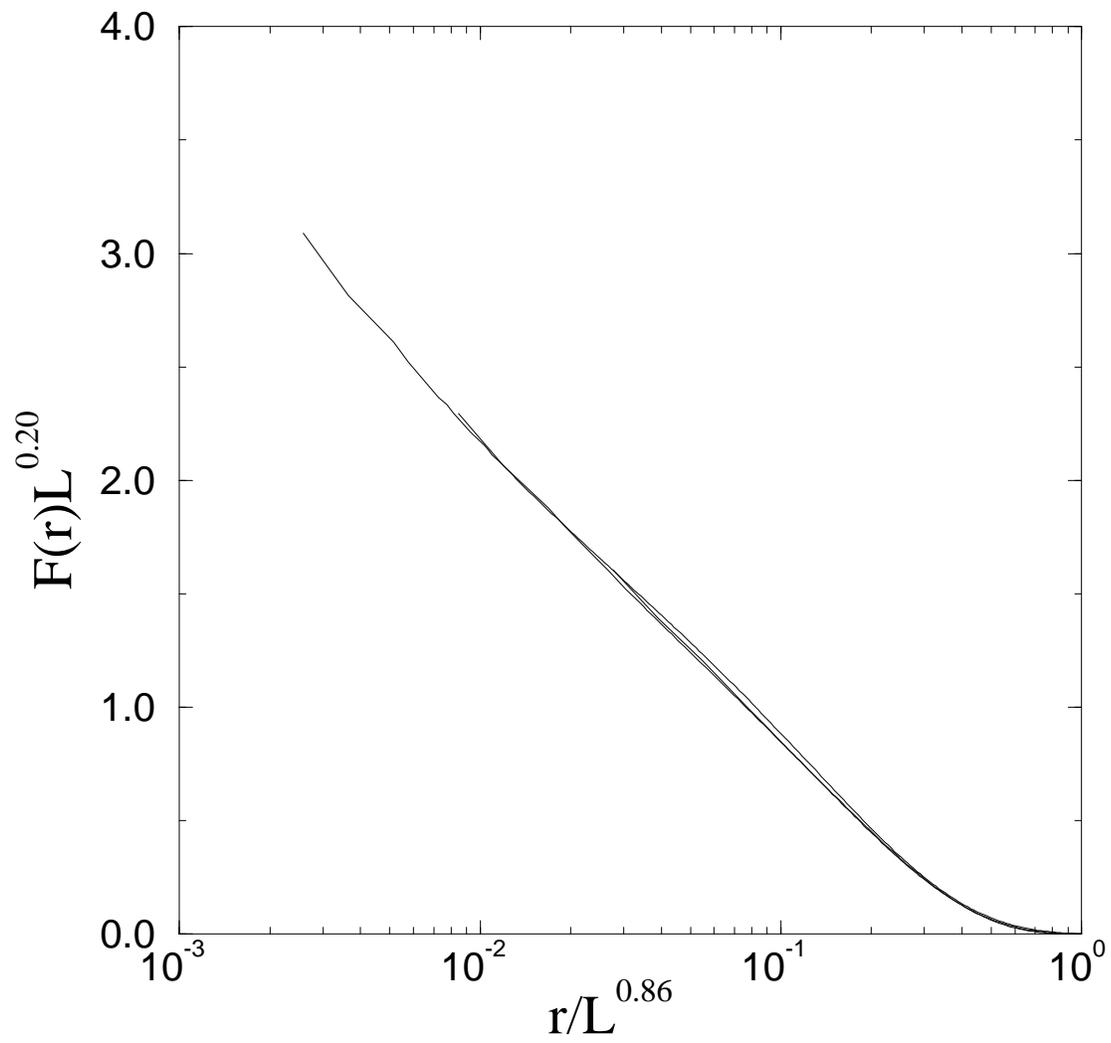
\begin{figure}
\caption{Scaling plot of the cumulative radial distribution
function $F(r)$. Plot of $F(r) L^{0.20}$ vs. $r/L^{0.86}$ shows
the data collapse for the system sizes $L= 65,
257$ and 1025.}
\end{figure}

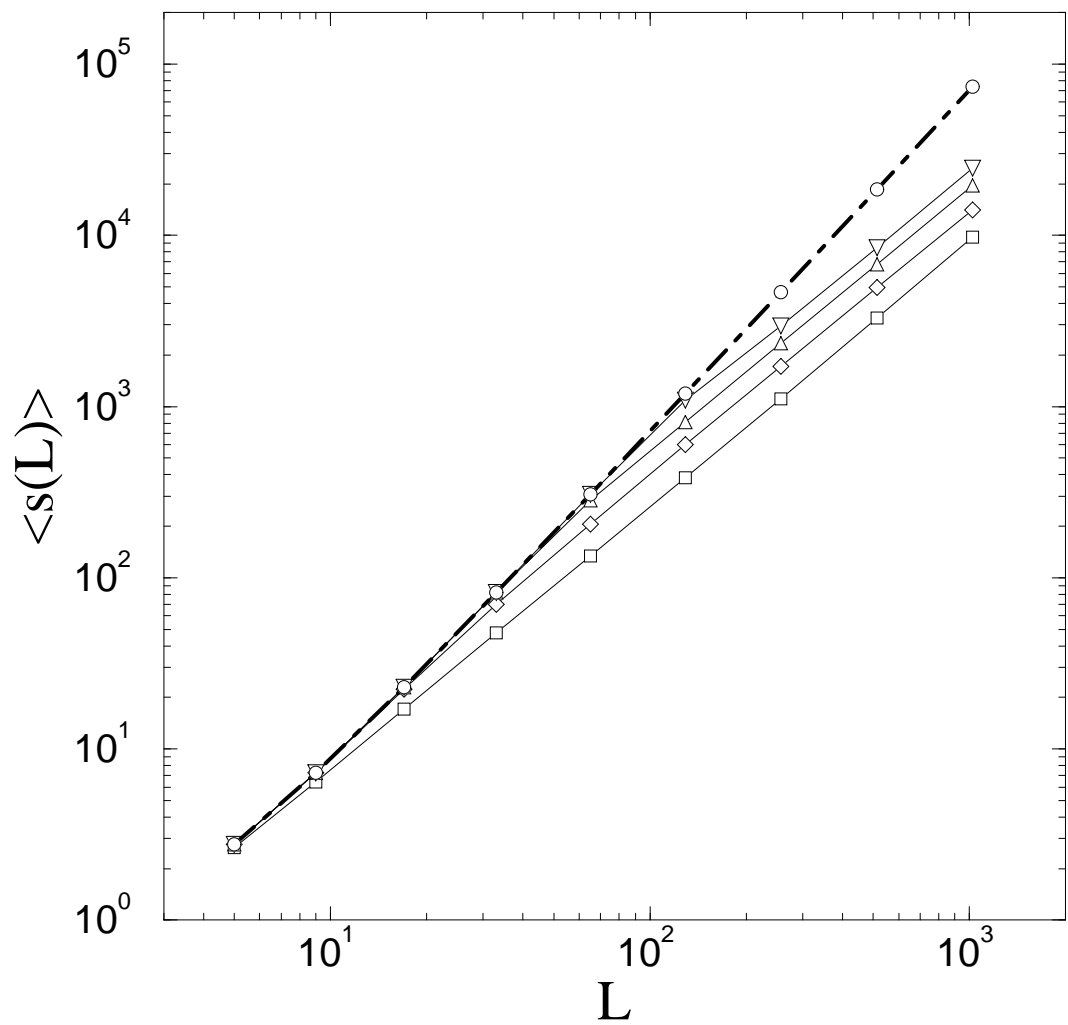
\begin{figure}
\caption{Plot of $<s(L)>$ versus $L$ for $m$ = 1, 2, 4 and 8
of SMAI (solid lines) and for ASM (dot dashed line). For
each value of $m$ there is a threshold system size
$L$ at which the cross-over from ASM behaviour to SMAI
takes place.}
\end{figure}

\newpage

\vskip 0.5 cm
{\centerline {\LARGE {\bf (a)}}}
\vskip 2.0 cm

\setlength{\unitlength}{0.00063300in}%
\begingroup\makeatletter\ifx\SetFigFont\undefined
% extract first six characters in \fmtname
\def\x#1#2#3#4#5#6#7\relax{\def\x{#1#2#3#4#5#6}}%
\expandafter\x\fmtname xxxxxx\relax \def\y{splain}%
\ifx\x\y   % LaTeX or SliTeX?
\gdef\SetFigFont#1#2#3{%
  \ifnum #1<17\tiny\else \ifnum #1<20\small\else
  \ifnum #1<24\normalsize\else \ifnum #1<29\large\else
  \ifnum #1<34\Large\else \ifnum #1<41\LARGE\else
     \huge\fi\fi\fi\fi\fi\fi
  \csname #3\endcsname}%
\else
\gdef\SetFigFont#1#2#3{\begingroup
  \count@#1\relax \ifnum 25<\count@\count@25\fi
  \def\x{\endgroup\@setsize\SetFigFont{#2pt}}%
  \expandafter\x
    \csname \romannumeral\the\count@ pt\expandafter\endcsname
    \csname @\romannumeral\the\count@ pt\endcsname
  \csname #3\endcsname}%
\fi
\fi\endgroup
\begin{picture}(8419,3624)(-1800,-4948)
\thicklines
\put(4085,-3333){\circle{642}}
\put(700,-2996){\circle{642}}
\put(2565,-2386){\framebox(1275,1050){}}
\put(615,-2386){\framebox(1275,1050){}}
\put(4515,-2386){\framebox(1275,1050){}}
\put(615,-1861){\line( 1, 0){1275}}
\put(2565,-1861){\line( 1, 0){1275}}
\put(4515,-1861){\line( 1, 0){1275}}
\put(5190,-1336){\line( 0,-1){1050}}
\put(465,-3286){\makebox(6.6667,10.0000){\SetFigFont{10}{12}{rm}.}}
\put(3165,-1336){\line( 0,-1){1050}}
\put(1290,-1336){\line( 0,-1){1050}}
\put(690,-4936){\framebox(1275,1050){}}
\put(1365,-3886){\line( 0,-1){1050}}
\put(690,-4411){\line( 1, 0){1275}}
\put(2640,-4936){\framebox(1275,1050){}}
\put(4590,-4936){\framebox(1275,1050){}}
\put(2640,-4411){\line( 1, 0){1275}}
\put(3240,-3886){\line( 0,-1){1050}}
\put(4590,-4411){\line( 1, 0){1275}}
\put(5265,-3886){\line( 0,-1){1050}}
\put(1965,-1861){\vector( 1, 0){525}}
\put(3915,-1861){\vector( 1, 0){525}}
\put(5940,-1861){\vector( 1, 0){525}}
\put( 90,-4411){\vector( 1, 0){525}}
\put(2040,-4411){\vector( 1, 0){525}}
\put(3990,-4411){\vector( 1, 0){525}}
\put(840,-2686){\vector( 1, 2){150}}
\put(3840,-3586){\vector(-1,-2){150}}
\put(840,-1786){\makebox(0,0)[lb]{\smash{\SetFigFont{29}{34.8}{bf}3}}}
\put(1515,-1786){\makebox(0,0)[lb]{\smash{\SetFigFont{29}{34.8}{bf}3}}}
\put(840,-2311){\makebox(0,0)[lb]{\smash{\SetFigFont{29}{34.8}{bf}3}}}
\put(1515,-2311){\makebox(0,0)[lb]{\smash{\SetFigFont{29}{34.8}{bf}0}}}
\put(2790,-1786){\makebox(0,0)[lb]{\smash{\SetFigFont{29}{34.8}{bf}3}}}
\put(3465,-2311){\makebox(0,0)[lb]{\smash{\SetFigFont{29}{34.8}{bf}0}}}
\put(4740,-1786){\makebox(0,0)[lb]{\smash{\SetFigFont{29}{34.8}{bf}4}}}
\put(5415,-2311){\makebox(0,0)[lb]{\smash{\SetFigFont{29}{34.8}{bf}1}}}
\put(915,-4336){\makebox(0,0)[lb]{\smash{\SetFigFont{29}{34.8}{bf}1}}}
\put(1590,-4336){\makebox(0,0)[lb]{\smash{\SetFigFont{29}{34.8}{bf}4}}}
\put(915,-4861){\makebox(0,0)[lb]{\smash{\SetFigFont{29}{34.8}{bf}0}}}
\put(1590,-4861){\makebox(0,0)[lb]{\smash{\SetFigFont{29}{34.8}{bf}1}}}
\put(2865,-4336){\makebox(0,0)[lb]{\smash{\SetFigFont{29}{34.8}{bf}1}}}
\put(3540,-4336){\makebox(0,0)[lb]{\smash{\SetFigFont{29}{34.8}{bf}1}}}
\put(2865,-4861){\makebox(0,0)[lb]{\smash{\SetFigFont{29}{34.8}{bf}0}}}
\put(3540,-4861){\makebox(0,0)[lb]{\smash{\SetFigFont{29}{34.8}{bf}2}}}
\put(4815,-4336){\makebox(0,0)[lb]{\smash{\SetFigFont{29}{34.8}{bf}1}}}
\put(5490,-4336){\makebox(0,0)[lb]{\smash{\SetFigFont{29}{34.8}{bf}2}}}
\put(4815,-4861){\makebox(0,0)[lb]{\smash{\SetFigFont{29}{34.8}{bf}0}}}
\put(5490,-4861){\makebox(0,0)[lb]{\smash{\SetFigFont{29}{34.8}{bf}2}}}
\put(3465,-1786){\makebox(0,0)[lb]{\smash{\SetFigFont{29}{34.8}{bf}3}}}
\put(2790,-2311){\makebox(0,0)[lb]{\smash{\SetFigFont{29}{34.8}{bf}4}}}
\put(5415,-1786){\makebox(0,0)[lb]{\smash{\SetFigFont{29}{34.8}{bf}3}}}
\put(4740,-2311){\makebox(0,0)[lb]{\smash{\SetFigFont{29}{34.8}{bf}0}}}
\put(615,-3136){\makebox(0,0)[lb]{\smash{\SetFigFont{29}{34.8}{bf}1}}}
\put(3990,-3511){\makebox(0,0)[lb]{\smash{\SetFigFont{29}{34.8}{bf}1}}}
\end{picture}
 
\vskip 2.0 cm
{\centerline {\LARGE {\bf (b)}}}
\vskip 1.0 cm

\setlength{\unitlength}{0.00063300in}%
\begingroup\makeatletter\ifx\SetFigFont\undefined
% extract first six characters in \fmtname
\def\x#1#2#3#4#5#6#7\relax{\def\x{#1#2#3#4#5#6}}%
\expandafter\x\fmtname xxxxxx\relax \def\y{splain}%
\ifx\x\y   % LaTeX or SliTeX?
\gdef\SetFigFont#1#2#3{%
  \ifnum #1<17\tiny\else \ifnum #1<20\small\else
  \ifnum #1<24\normalsize\else \ifnum #1<29\large\else
  \ifnum #1<34\Large\else \ifnum #1<41\LARGE\else
     \huge\fi\fi\fi\fi\fi\fi
  \csname #3\endcsname}%
\else
\gdef\SetFigFont#1#2#3{\begingroup
  \count@#1\relax \ifnum 25<\count@\count@25\fi
  \def\x{\endgroup\@setsize\SetFigFont{#2pt}}%
  \expandafter\x
    \csname \romannumeral\the\count@ pt\expandafter\endcsname
    \csname @\romannumeral\the\count@ pt\endcsname
  \csname #3\endcsname}%
\fi
\fi\endgroup
\begin{picture}(8669,4545)(-1800,-4883)
\thicklines
\put(2115,-698){\circle{692}}
\put(2640,-4523){\circle{692}}
\put(2565,-2386){\framebox(1275,1050){}}
\put(4515,-2386){\framebox(1275,1050){}}
\put(615,-1861){\line( 1, 0){1275}}
\put(2565,-1861){\line( 1, 0){1275}}
\put(4515,-1861){\line( 1, 0){1275}}
\put(2565,-3811){\framebox(1275,1050){}}
\put(5115,-2761){\line( 0,-1){1050}}
\put(4515,-3286){\line( 1, 0){1275}}
\put(2565,-3286){\line( 1, 0){1275}}
\put(4515,-3811){\framebox(1275,1050){}}
\put(465,-3286){\makebox(6.6667,10.0000){\SetFigFont{10}{12}{rm}.}}
\put(8715,-2236){\makebox(6.6667,10.0000){\SetFigFont{10}{12}{rm}.}}
\put(8715,-2236){\makebox(6.6667,10.0000){\SetFigFont{10}{12}{rm}.}}
\put(1965,-1861){\vector( 1, 0){525}}
\put(3915,-1861){\vector( 1, 0){525}}
\put(3915,-3286){\vector( 1, 0){525}}
\put( 90,-3286){\vector( 1, 0){525}}
\put(5865,-1861){\vector( 1, 0){525}}
\put(690,-3286){\line( 1, 0){1275}}
\put(690,-3811){\framebox(1275,1050){}}
\put(2040,-3286){\vector( 1, 0){450}}
\put(5190,-1336){\line( 0,-1){1050}}
\put(3240,-1336){\line( 0,-1){1050}}
\put(615,-2386){\framebox(1275,1050){}}
\put(1215,-1336){\line( 0,-1){1050}}
\put(1290,-2761){\line( 0,-1){1050}}
\put(3240,-2761){\line( 0,-1){1050}}
\put(1876,-961){\vector(-1,-3){127.500}}
\put(2776,-4186){\vector( 2, 3){242.308}}
\put(840,-1786){\makebox(0,0)[lb]{\smash{\SetFigFont{29}{34.8}{bf}3}}}
\put(1515,-1786){\makebox(0,0)[lb]{\smash{\SetFigFont{29}{34.8}{bf}3}}}
\put(840,-2311){\makebox(0,0)[lb]{\smash{\SetFigFont{29}{34.8}{bf}3}}}
\put(1515,-2311){\makebox(0,0)[lb]{\smash{\SetFigFont{29}{34.8}{bf}0}}}
\put(2790,-1786){\makebox(0,0)[lb]{\smash{\SetFigFont{29}{34.8}{bf}3}}}
\put(3465,-1786){\makebox(0,0)[lb]{\smash{\SetFigFont{29}{34.8}{bf}4}}}
\put(2790,-2311){\makebox(0,0)[lb]{\smash{\SetFigFont{29}{34.8}{bf}3}}}
\put(3465,-2311){\makebox(0,0)[lb]{\smash{\SetFigFont{29}{34.8}{bf}0}}}
\put(5415,-1786){\makebox(0,0)[lb]{\smash{\SetFigFont{29}{34.8}{bf}0}}}
\put(5415,-2311){\makebox(0,0)[lb]{\smash{\SetFigFont{29}{34.8}{bf}1}}}
\put(840,-3211){\makebox(0,0)[lb]{\smash{\SetFigFont{29}{34.8}{bf}1}}}
\put(1515,-3211){\makebox(0,0)[lb]{\smash{\SetFigFont{29}{34.8}{bf}0}}}
\put(840,-3736){\makebox(0,0)[lb]{\smash{\SetFigFont{29}{34.8}{bf}4}}}
\put(1515,-3736){\makebox(0,0)[lb]{\smash{\SetFigFont{29}{34.8}{bf}1}}}
\put(2790,-3211){\makebox(0,0)[lb]{\smash{\SetFigFont{29}{34.8}{bf}1}}}
\put(3465,-3211){\makebox(0,0)[lb]{\smash{\SetFigFont{29}{34.8}{bf}0}}}
\put(2790,-3736){\makebox(0,0)[lb]{\smash{\SetFigFont{29}{34.8}{bf}1}}}
\put(4740,-3211){\makebox(0,0)[lb]{\smash{\SetFigFont{29}{34.8}{bf}1}}}
\put(5415,-3211){\makebox(0,0)[lb]{\smash{\SetFigFont{29}{34.8}{bf}0}}}
\put(4740,-3736){\makebox(0,0)[lb]{\smash{\SetFigFont{29}{34.8}{bf}2}}}
\put(5415,-3736){\makebox(0,0)[lb]{\smash{\SetFigFont{29}{34.8}{bf}2}}}
\put(3465,-3736){\makebox(0,0)[lb]{\smash{\SetFigFont{29}{34.8}{bf}2}}}
\put(2565,-4636){\makebox(0,0)[lb]{\smash{\SetFigFont{29}{34.8}{bf}1}}}
\put(2040,-811){\makebox(0,0)[lb]{\smash{\SetFigFont{29}{34.8}{bf}1}}}
\put(4740,-1786){\makebox(0,0)[lb]{\smash{\SetFigFont{29}{34.8}{bf}4}}}
\put(4815,-2311){\makebox(0,0)[lb]{\smash{\SetFigFont{29}{34.8}{bf}3}}}
\end{picture}
\vfill
\small
\leftline {Figure 1 : Manna and Giri}
\end {document}